\documentclass[aps,prl,preprint,superscriptaddress]{revtex4-1}
\usepackage{amsmath}
\usepackage{graphicx}
\begin{document}

%\preprint{}

\title{Recoil Polarization Measurements of the Proton Electromagnetic
  Form Factor Ratio to $Q^2 = 8.5$ GeV$^2$}


\author{A. J. R. Puckett} \email[Corresponding author: ]{puckett@jlab.org}
\affiliation{Massachusetts Institute of Technology, Cambridge, MA
  02139}
\author{E. J. Brash}
\affiliation{Christopher Newport University, Newport News, VA 23606}
\affiliation{Thomas Jefferson National Accelerator Facility, Newport News, VA 23606}
\author{M. K. Jones}
\affiliation{Thomas Jefferson National Accelerator Facility, Newport News, VA 23606}
\author{W. Luo}
\affiliation{Lanzhou University, Lanzhou 730000, Gansu, Peoples Republic of China}
\author{M. Meziane}
\author{L. Pentchev}
\author{C. F. Perdrisat}
\affiliation{College of William and Mary, Williamsburg, VA 23187}
\author{V. Punjabi}
\author{F. R. Wesselmann}
\affiliation{Norfolk State University, Norfolk, VA 23504}
%\thanks{}
\author{A. Ahmidouch}
\affiliation{North Carolina A\&T State University, Greensboro, NC 27411}
% ...

\date{\today}

\begin{abstract}
Among the most fundamental observables of
nucleon structure, electromagnetic form factors are a crucial benchmark for modern  
calculations describing the strong interaction dynamics of the
nucleon's quark constituents; indeed, recent proton
data have attracted intense theoretical interest.
In this letter, we report new
measurements of the proton electromagnetic form factor ratio using the
recoil polarization method, at momentum transfers  
$Q^2=5.2$, 6.7, and 8.5 GeV$^2$.  By extending the range of $Q^2$ for which  
$G_E^p$ is accurately determined by more than 50\%, these measurements
will provide  
significant constraints on models of nucleon structure in  
the non-perturbative regime.
\end{abstract}

%\keywords{}

\maketitle

% Points to make in introduction:
% Intense investigation, current interest
% recoil polarization superior to Rosenbluth
% GM dominance of cross section
% connection to charge density?
% form factor definitions: Sachs vs. Dirac/Pauli

\paragraph{}
The measurement of nucleon electromagnetic form factors, pioneered at
Stanford in the 1950s, has again become the subject of intense
investigation. Precise recoil polarization experiments
\cite{Jones00,*Punjabi05,*Gayou02} established conclusively that the proton
electric form factor $G_E^p$ falls faster than the magnetic form
factor $G_M^p$ for momentum transfers $Q^2 \ge 1$ GeV$^2$, in
disagreement with results obtained from cross section measurements \cite{PerdrisatPunjabiVanderhaegen2007,Andi94,Christy04,Qattan05}. Precise
data to the highest possible $Q^2$ are
needed, for example, to test the onset of validity of
perturbative QCD (pQCD) predictions for asymptotic form factor
behavior \cite{BrodskyLepage1979}, constrain Generalized
Parton Distributions (GPDs) \cite{JiGPDPRL1997}, and to determine the
nucleon's model-independent impact parameter-space charge and
magnetization densities \cite{MillerGPDchargedensity2007,*MillerGPDmagnetdensity2008}.

The effect of nucleon structure on
elastic electron-nucleon scattering at a spacelike momentum transfer $q^2 = -Q^2
< 0$ is described in the
one-photon-exchange approximation by the helicity-conserving and
helicity-flip form factors $F_1(q^2)$ (Dirac) and $F_2(q^2)$
(Pauli), or alternatively the Sachs form factors, defined as the linear combinations
$G_E = F_1 - \tau F_2$ (electric) and $G_M = F_1 + F_2$ (magnetic),
where $\tau \equiv Q^2 / 4M^2$ and $M$ is the nucleon mass. Polarization observables, such as the beam-target double-spin
asymmetry \cite{Dombey1969} and polarization transfer \cite{AkhiezerRekalo2,ArnoldCarlsonGross} provide enhanced sensitivity to the electric
form factor at large $Q^2$ compared to cross section measurements, for
which $G_M$ becomes the dominant contribution. The polarization of the
recoil proton in the elastic scattering of longitudinally polarized
electrons from unpolarized protons has longitudinal ($P_l$) and transverse ($P_t$) components with respect to
the momentum transfer in the scattering plane \cite{ArnoldCarlsonGross}. The ratio $P_t/P_l$ is
proportional to $G_E^p/G_M^p$:
\begin{eqnarray}
  R \equiv \mu_p \frac{G_E^p}{G_M^p} &=& -\mu_p \frac{P_t}{P_l} \frac{E_e+E'_e}{2M_p} \tan
  \frac{\theta_e}{2} \label{FFratioformula} 
\end{eqnarray}
where $\mu_p$ is the proton magnetic moment, $E_e$ is the beam energy, $E'_e$ is the scattered $e^-$ energy,
$\theta_e$ is the $e^-$ scattering angle and $M_p$ is the proton
mass. Because the extraction of $G_E^p$ from the ratio
\eqref{FFratioformula} is much less sensitive than the Rosenbluth
method \cite{Rosenbluth1950} to higher-order corrections beyond the
standard radiative corrections \cite{AfanasevRadCorr}, it
is generally believed that polarization measurements
provide the correct determination of $G_E^p$ in the $Q^2$ range where
the two methods disagree. Previously neglected two-photon-exchange effects have been
shown to partially resolve the discrepancy \cite{TPEXreview},
and are a highly active area of theoretical and experimental
investigation.

The new measurements of $G_E^p/G_M^p$ were carried out in experimental Hall C at
Jefferson Lab. A continuous polarized electron beam was scattered from
a 20 cm liquid
hydrogen target, and elastically scattered electrons and protons were
detected in coincidence. Typical beam currents ranged from 60-100
$\mu$A. The beam helicity was reversed
pseudorandomly at
30 Hz. The beam polarization of typically 80-85\% was
monitored periodically using M\"{o}ller polarimetry \cite{MollerNIM}.
% The word "accuracy" implies systematic/total error

Scattered protons were detected in the Hall C High
Momentum Spectrometer (HMS) \cite{HornLongPaper}, a superconducting magnetic spectrometer
with three focusing quadrupole magnets followed by a $25^\circ$ vertical bend dipole
magnet, operated in a point-to-point tune. Charged particle trajectories at the focal plane were measured
using drift chambers, and their momenta, scattering angles, and vertex
coordinates were reconstructed using the transport matrix of the HMS. For this experiment, the HMS
trigger was defined by a coincidence between the pair of scintillator
planes just behind the drift chambers and an additional scintillator paddle
placed at the exit of the dipole. The size of this new paddle matched
the acceptance of elastically scattered protons.

To measure the polarization of scattered protons, a
double Focal Plane Polarimeter (FPP) was installed in the HMS detector
hut, replacing the standard Cerenkov detector and rear scintillators. The FPP consists of two retractable 50 g cm$^{-2}$ CH$_2$ analyzer doors, each
followed by a pair of large-acceptance drift chambers with an active
area $164 \times 132$ cm$^2$. The tracks of protons scattered in the
analyzer material were reconstructed with an angular resolution of approximately 1
mrad.

Scattered electrons were detected in a large-acceptance electromagnetic calorimeter
(BigCal) positioned for each $Q^2$ to cover a solid angle
kinematically matched to the $\approx 7$ msr proton acceptance of the
HMS, up to 143 msr at $Q^2 = 8.5$ GeV$^2$.
BigCal was assembled from 1,744 lead-glass bars stacked in a
rectangular array with a frontal area of $1.2 \times 2.2$ m$^2$ and a
thickness of approximately 15 radiation lengths. The trigger for BigCal
was formed from analog sums of up to 64 channels, grouped with overlap
to maximize the efficiency for electrons at high thresholds of nearly half
the elastic $e^-$ energy, used to suppress charged pions and low-energy
backgrounds. The over-determined elastic $ep$ kinematics allowed for continuous \emph{in
  situ} calibration and gain matching. The primary trigger for the
experiment was a time coincidence
between BigCal and the HMS within a $\pm$50 ns window. 

% Table \ref{kintable} shows the central
% kinematic settings for each $Q^2$ point.
% \begin{table}[h]
%   \begin{center}
%     \begin{tabular}{|ccccccc|}
%       \hline 
%       $Q^2$, GeV$^2$ & $p_p$, GeV & $\theta_p$, $^\circ$ & $E_e$, GeV &
%       $E'_e$, GeV & $\theta_e$, $^\circ$ & $R_{cal}$, m \\ \hline 
%       5.2 & 3.589 & 17.9 & 4.05 & 1.27 & 60.3 & 6.05 \\ 
%       6.8 & 4.464 & 19.1 & 5.71 & 2.09 & 44.2 & 6.08 \\
%       8.5 & 5.407 & 11.6 & 5.71 & 1.16 & 69.0 & 4.30 \\ \hline
%     \end{tabular}
%   \end{center}
%   \caption{\label{kintable} Central kinematic settings. $R_{cal}$ is
%     the distance from the nominal origin of Hall C to the surface of BigCal.}
% \end{table}

Elastic events were selected by applying cuts to enforce two-body reaction
kinematics. The electron scattering angle $\theta_e$ was
predicted from the proton momentum $p_p$ and the beam energy, and the
azimuthal angle $\phi_e$ was predicted from $\phi_p$ assuming
coplanarity of the electron and the proton. The predicted electron trajectory was projected
from the interaction vertex to the surface of BigCal and compared to
the measured shower coordinates. The small area of
each cell relative to the transverse shower size resulted
in coordinate resolution of 5-10 mm, corresponding to an
angular resolution of 1-3 mrad, which matched or exceeded the
resolution of the predicted angles from elastic
kinematics of the reconstructed proton. 

% This cut effectively suppresses the initial-state
% Bremsstrahlung component of the elastic radiative tail, which causes a positive
% displacement along the $\Delta x$ axis.

\begin{figure}[h]
  \begin{center}
    \includegraphics[angle=90,width=0.46\textwidth]{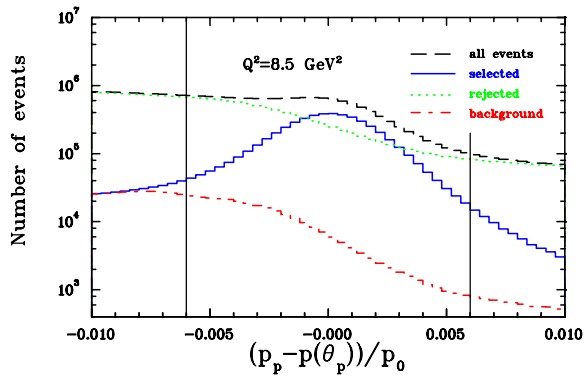}
  \end{center}
  \caption{\label{pmisspfig} (Color online) Elastic event selection for $Q^2=8.5$
    GeV$^2$. The momentum difference $(p_p-p_p(\theta_p))/p_0$, where
    $p_0$ is the HMS central momentum, plotted for all events (black
    dashed), events passing the $3\sigma$ elliptical cut (blue solid), and events
    failing the cut (green dotted). The estimated background (red
    dot-dashed) integrated over the final cut region (black vertical
    lines) is approximately 5.9\%.}
\end{figure}

An elliptical cut $\left(\Delta x/x_{max}\right)^2
  + \left(\Delta y/y_{max}\right)^2 \le 1$ was applied to the
horizontal and vertical coordinate differences $(\Delta x, \Delta
y)$, where $(x_{max}, y_{max})$ are the $Q^2$-dependent, $3\sigma$ cut widths used for
the final analysis. An additional cut was applied to the proton
angle-momentum correlation $p_p - p_p(\theta_p)$ which further suppressed
the inelastic background. No cut was applied to the measured $e^-$
energy, because the BigCal energy resolution was insufficient to provide additional separation between elastic and
inelastic events. Figure
\ref{pmisspfig} illustrates the separation of the elastic peak
in the $p_p-p_p(\theta_p)$ spectrum using BigCal.

The dominant background was hard-Bremsstrahlung-induced
$\pi^0$ photoproduction, $\gamma + p \rightarrow \pi^0 + p$, in the
2.3\% radiation length cryotarget, with the proton detected in the HMS
and one or two
$\pi^0$ decay photons detected in BigCal. The kinematics of this
reaction overlap with elastic $ep$ scattering within experimental
resolution for near-endpoint photons. The
contribution of quasi-elastic Al$(e,e'p)$ scattering from the cryocell
windows was also measured and found to be
negligible after cuts. The total background including inelastic
reactions and
random coincidences was
estimated as a function of $p_p-p_p(\theta_p)$, as shown in figure \ref{pmisspfig}, using a two-dimensional Gaussian extrapolation of the
$(\Delta x, \Delta y)$ distribution of the background into the cut
region under the
elastic peak. A Monte Carlo simulation of elastic $ep$ scattering and
$\pi^0$ photoproduction was performed as a check on the background
estimation procedure. The two methods agreed at the 10\%
(relative) level for wide variations of the cuts.

% As a check on the first method, the
% standard Hall C Monte Carlo code SIMC was used to estimate the relative contributions of
% elastic $ep$ scattering and $\pi^0$ photoproduction within the
% experimental acceptance. Agreement between the two methods at the
% $\pm10$\% level (relative) was stable with respect to independent
% variations of the $\Delta x$, $\Delta y$, and $p-p(\theta_p)$ cut widths between 3$\sigma$ and 6$\sigma$. $3\sigma$
% cuts were used for the final analysis.

\begin{figure}[h]
  \begin{center}
    \includegraphics[angle=90,width=.46\textwidth]{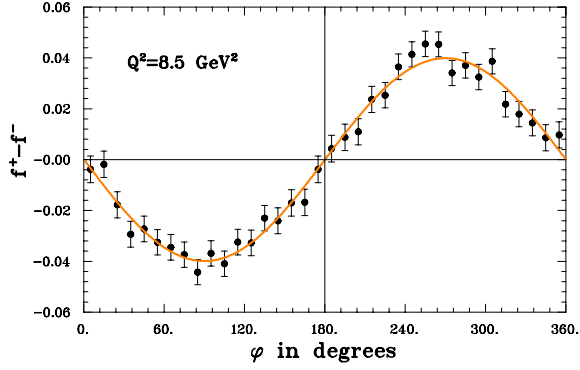}
  \end{center}
  \caption{\label{asymfig} (Color online) Helicity difference distribution $f_+ -
    f_-$ for $Q^2=8.5$ GeV$^2$, $0.5^\circ \le \vartheta \le
    14.0^\circ$. The data are fitted with $f_+ - f_- = a \cos \varphi
    + b \sin \varphi$ (solid curve), resulting in $a = (0.16 \pm 1.19)
    \times 10^{-3}$ and $b = (-3.99 \pm 0.12) \times 10^{-2}$
    ($\chi^2/\mbox{n.d.f.} = 0.67$).}
\end{figure}

The angular
distribution of protons scattered in the CH$_2$ analyzers measures the
polarization components at the focal plane. The
polar and azimuthal scattering angles $(\vartheta,\varphi)$ of tracks in the FPP drift chambers
were calculated relative to the incident track defined by the focal
plane drift chambers. The measured angular distribution can be
expressed in the general form,
\begin{eqnarray}
  N^\pm(p,\vartheta,\varphi) &=& N_0^\pm
  \frac{\varepsilon(p,\vartheta)}{2\pi}\Big[1+(c_1\pm A_y P_y^{fpp})\cos \varphi + \nonumber \\
    & & (s_1 \mp A_yP_x^{fpp})\sin \varphi + \nonumber \\
    & & c_2 \cos(2\varphi) + s_2 \sin(2\varphi) + \ldots \Big] \label{angulardistribution}
\end{eqnarray}
where $N_0^{\pm}$ is the number of incident protons in the $\pm$ beam
helicity state, $\varepsilon(p,\vartheta)$ is the fraction of
protons of momentum $p$ scattered by an angle $\vartheta$, $A_y(p,\vartheta)$ is the analyzing power of the $\vec{p}+$CH$_2$
reaction, and $P_x^{fpp}$ and $P_y^{fpp}$ are the transverse
components of the proton polarization at the focal plane. $c_1, s_1,
c_2, s_2, \ldots$ are the Fourier coefficients of helicity-independent
instrumental asymmetries, which are cancelled to first order by the
helicity reversal. 
Figure \ref{asymfig} shows the measured
helicity-dependent azimuthal asymmetry $f_+ - f_- = \frac{2\pi}{\Delta
  \varphi}\left[\frac{N_+(\varphi)}{N_0^+}-\frac{N_-(\varphi)}{N_0^-}\right]
\approx \bar{A}_y \left[P_y^{fpp} \cos \varphi - P_x^{fpp} \sin
  \varphi \right]$, where $\Delta \varphi$ is the bin
width, summed over all $p$ and the $\vartheta$ range $0.5^\circ \le
\vartheta \le 14^\circ$ outside which $A_y \approx 0$. 
%The phase shift of the asymmetry expressed in the form $f_+ -
%f_- = A \sin ( \varphi + \delta)$ measures the
%ratio $P_y^{fpp} / P_x^{fpp} = \tan \delta$.

The extraction of $P_t$, $P_l$, and $P_t/P_l$ from the measured
asymmetry at the focal plane involves the precession of
the proton polarization in the HMS magnetic field, governed by the Thomas-BMT equation \cite{BMTequation}. The
rotation of longitudinal $P_l$ into normal $P_x^{fpp}$ allows the
simultaneous measurement of $P_t$ and $P_l$ in the FPP,
which is insensitive to longitudinal polarization. The unique spin
transport matrix for each proton trajectory was calculated as a function
of its angles, momentum, and vertex coordinates from a detailed model of the HMS using the
differential-algebra based COSY software \cite{COSY}. The polarization components
at the target were then extracted by maximizing the likelihood function defined as:
\begin{eqnarray}
  \mathcal{L}(P_t, P_l) &=&  \prod_{i=1}^{N_{event}} \Big[ 1 + h \epsilon_i A_y^{(i)} ( S_{yt}^{(i)} P_t + S_{yl}^{(i)}P_l ) \cos \varphi_i \nonumber \\
  & & - h \epsilon_i A_y^{(i)} ( S_{xt}^{(i)} P_t + S_{xl}^{(i)} P_l ) \sin \varphi_i + \lambda_0^{(i)} \Big] 
\end{eqnarray}
where $h$ is the beam polarization, $S_{jk}^{(i)}$ are the spin
transport matrix elements, $\epsilon_i=\pm1$ is the beam helicity, and
$\lambda_0$ is the false asymmetry. 
% For a simple dipole, $S_{yt} = 1$,
% $S_{yl} = S_{xt} = 0$, and $S_{xl} = -\sin \chi_\theta$, where
% $\chi_\theta = \gamma \kappa_p \theta_{bend}$ is the product of the
% proton's boost factor, anomalous magnetic moment and trajectory bend
% angle in the dipole. 

The polarization of the residual inelastic
background passing ``elasticity'' cuts was obtained from
the rejected events using the same procedure, and used to
correct the polarization of elastic events. The acceptance-averaged fractional inelastic
backgrounds for $Q^2=5.2$, $6.7$, and
$8.5$ GeV$^2$ were $N_{inel}/(N_{inel}+N_{el}) = (1.12
\pm 0.16) \%$, $(0.77 \pm 0.12 )\%$, and $(5.9\pm 0.9)\%$,
respectively. The resulting 
absolute corrections to $R$ were $\Delta R = (8.4 \pm 1.5) \times 10^{-3}$, $(7.5\pm1.3) \times
10^{-3}$, and $(6.0 \pm 1.3) \times 10^{-2}$.

Since the beam polarization and the $\vec{p}+$CH$_2$ analyzing power cancel in the ratio, there are few significant sources of systematic
uncertainty in the results of this experiment. The most important
contribution comes from the precession calculation. An excellent approximation to
the full COSY calculation used for the final analysis is obtained from
the product
of simple rotations relative to the proton trajectory by angles
$\chi_\phi$ in the non-dispersive plane and $\chi_\theta$ in the
dispersive plane. $\chi_\phi = \gamma \kappa_p \phi_{bend}$
and $\chi_\theta = \gamma \kappa_p \theta_{bend}$ are proportional
to the trajectory bend angles $\phi_{bend}$ and $\theta_{bend}$ by a factor equal to the product of the proton's boost factor $\gamma$
and anomalous magnetic moment $\kappa_p$. The relevant matrix
elements in this approximation are $S_{yt} = \cos \chi_\phi$, $S_{yl} = \sin \chi_\phi$,
$S_{xt} = \sin \chi_\phi \sin \chi_\theta$, and $S_{xl} = -\cos
\chi_\phi \sin \chi_\theta$. These simple matrix elements were used to
study the effects of systematic errors in the
reconstructed kinematics.

The error $\Delta \phi_{bend}$ due to unknown
misalignments of the quadrupoles relative to the HMS optical axis
leads to an error $\gamma \kappa_p \Delta \phi_{bend}$ on $P_t/P_l$. 
This uncertainty was minimized through a dedicated study of the
non-dispersive optics of the
HMS following the method of \cite{QuadAlign}, setting a conservative
upper limit of $\left|\Delta \phi \right| \le 0.5$ mrad, which is the single largest
contribution to the systematic uncertainty in
$R$. The contribution of uncertainties in the absolute
central momentum of the HMS and the dispersive bend angle
$\theta_{bend}$ is small by comparison. The extracted form factor ratio showed no statistically significant dependence on any of the variables involved in the
precession calculation, providing a strong test of its quality.

Uncertainties in $E_e$, $E'_e$ and $\theta_e$ make an even smaller
contribution. Uncertainties in the scattering angles in the FPP were
minimized by a software alignment procedure using
``straight-through'' data obtained with the CH$_2$ doors open. False
asymmetry coefficients obtained from Fourier analysis of
the helicity sum distribution $f_+ + f_-$ were used to correct the
small, second-order contributions to the extracted
polarization components. The resulting
correction to $R$ was small ($|\Delta R| \le 0.007$) and
negative for each $Q^2$. The correction procedure was verified using a Monte Carlo simulation. 

% The
% uncertainties are statistics-limited, as the systematic
% uncertainties, which include all the contributions discussed in the
% text combined in quadrature, are at most $~25\%$ of the statistical
% errors.
\begin{table}[h]
  \begin{center}
    \begin{tabular}{|cccc|}
      % columns: avg. Q2 (weighted by Ay^2?), Q2min, Q2max, muGE/GM +- dstat +- dsyst. 
      \hline $E_{e}$, GeV & $\theta_e$, $^\circ$ & $\left<Q^2\right>
      \pm \Delta Q^2$, GeV$^2$ & $R \pm \Delta R_{stat.} \pm \Delta R_{syst.}$ \\ \hline \hline
      4.05 & 60.3 & 5.17 $\pm$ 0.123 & $0.443 \pm 0.066 \pm 0.018$ \\ 
      5.71 & 44.2 & 6.70 $\pm$ 0.190 & $0.327 \pm 0.105 \pm 0.022$ \\
      5.71 & 69.0 & 8.49 $\pm$ 0.167 & $0.138 \pm 0.179 \pm 0.043$ \\ \hline
    \end{tabular}
    \caption{\label{resultstable} Results for $R = \mu_p
      G_E^p/G_M^p$, with statistical and systematic
      uncertainties. $E_e$ is the beam energy, $\theta_e$ is the
      central electron scattering angle, $\left<Q^2\right>$ is the acceptance-averaged
      $Q^2$, and $\Delta Q^2$ is the r.m.s. $Q^2$ acceptance.}
  \end{center}
\end{table}

\begin{figure}[h]
  \begin{center}
    \includegraphics[width=0.46\textwidth]{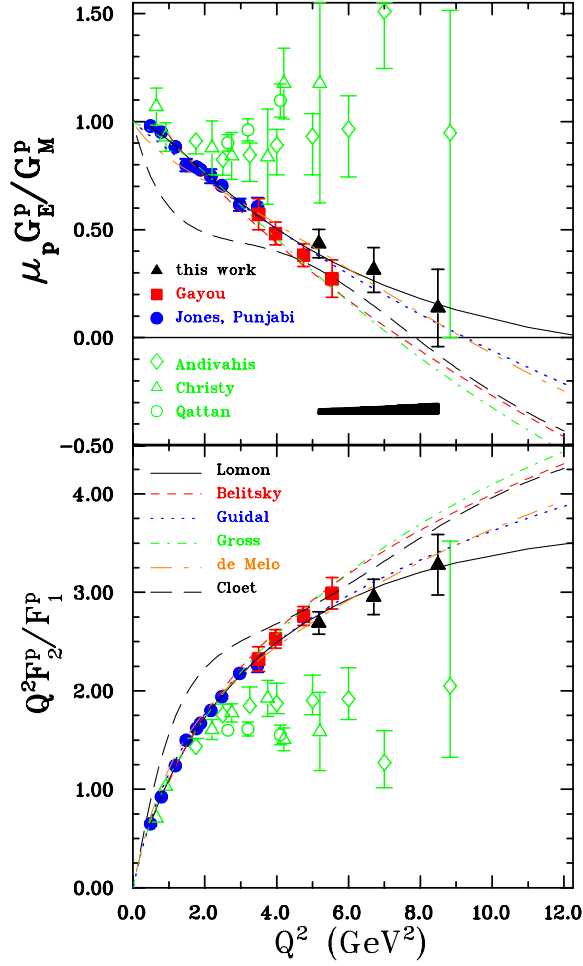}
  \end{center}
  \caption{\label{ratiofig} (Color online) Upper panel: The proton form factor ratio
    $\mu_pG_E^p/G_M^p$ from this experiment (filled black triangles),
    with statistical error bars and systematic error band below the data. Previous
    experiments are \cite{Punjabi05} (Jones, Punjabi, Gayou), 
    \cite{Andi94} (Andivahis), \cite{Christy04} (Christy), and \cite{Qattan05}
    (Qattan). Theory curves are
    \cite{Lomon2002, *Lomon2006} (Lomon), \cite{deMelo09} (de Melo),
    \cite{Gross06,*Gross08} (Gross), \cite{Cloet09}
    (Clo\"{e}t), \cite{GuidalGPD2005} (Guidal), and \cite{BelitskyJiYuan2003}
    (Belitsky). Lower panel: The same data and theory curves as
    the upper panel, expressed as $Q^2F_2^p / F_1^p$.}
\end{figure}

The results of the experiment are presented in table
\ref{resultstable}. Standard radiative corrections to $P_t/P_l$ were
calculated using the code MASCARAD \cite{AfanasevRadCorr}, found to
be no greater than $0.13\%$ (relative) for any of the three $Q^2$
values, and were not applied. Figure \ref{ratiofig} presents the new
results with
recent Rosenbluth and polarization data and selected theoretical
predictions.

Theoretical descriptions of nucleon form factors emphasize the
importance of both baryon-meson
and quark-gluon dynamics, with the former (latter) generally
presumed to dominate in the low (high) energy limit. Recent Vector Meson
Dominance (VMD) model fits by Lomon \cite{Lomon2002,*Lomon2006}
include $\rho'(1450)$ and $\omega'(1420)$ mesons in addition to the
usual $\rho$, $\omega$, and $\phi$, and a ``direct coupling'' term
enforcing pQCD-like behavior as $Q^2 \rightarrow \infty$. de Melo et
al. \cite{deMelo09} considered the
non-valence components of the nucleon state in a light-front
framework, using Ans\"atze for the
nucleon Bethe-Salpeter amplitude and a microscopic version of the VMD
model. Gross and Agbakpe \cite{Gross06,*Gross08} modeled the nucleon
as a bound state of three
dressed valence constituent quarks in a covariant spectator
theory. Clo\"et et al. \cite{Cloet09} calculated a dressed-quark core contribution to the
nucleon form factors in an approach based on Dyson-Schwinger equations
(DSE) in QCD. The disagreement between this calculation and the data
at lower $Q^2$ is attributed to the omission of meson cloud effects. 

The Dirac and Pauli form factors are related to the vector ($H$) and
tensor ($E$) GPDs through sum rules \cite{JiGPDPRL1997}. Guidal et al. \cite{GuidalGPD2005} fit a model of
the valence quark GPDs based on Regge phenomenology to form factor
data. In this model, the ratio $F_2^p/F_1^p$ constrains the $x
\rightarrow 1$ behavior of $E$, where $x$ is the light-cone parton
momentum fraction. When combined with the forward limit
of $H$ determined by parton distribution functions, the new
information on $E$ obtained from precise form factor data allowed an evaluation
of Ji's sum rule \cite{JiGPDPRL1997} for the total angular momentum
carried by quarks in the nucleon.

The data do not yet satisfy the leading-twist, leading order pQCD
``dimensional scaling'' relation $F_2^p \propto F_1^p/Q^2$
\cite{BrodskyLepage1979}. The modified scaling $Q^2 F_2^p/F_1^p \propto
\ln^2(Q^2/\Lambda^2)$ obtained by considering the subleading twist
components of the light-cone nucleon wavefunction \cite{BelitskyJiYuan2003}, with $\Lambda = 300$ MeV
as shown in figure \ref{ratiofig}, describes
the polarization data rather well. This
``precocious scaling'' of $F_2^p/F_1^p$ is a necessary, but not sufficient
condition for the validity of a pQCD description of nucleon form
factors. Despite progress in calculations based on light
cone QCD sum rules \cite{LCQCD2006}, pQCD form factor predictions have not yet reached the level of accuracy of phenomenological models such as
\cite{Lomon2002,Gross06,deMelo09,GuidalGPD2005} when applied to all
four form factors ($F_{1,2}^{p,n}$), underscoring both the
difficulty of predicting observables of hard exclusive reactions directly from QCD, and the strong
guidance to theory provided by high quality data such as the results
reported in this letter.

% In this context, the new data favor a larger value of
% the diquark radius (0.8 fm for the curve shown in figure \ref{ratiofig}),
% the sole model parameter related to the position of the zero of
% $G_E^p$. The disagreement between this calculation and the data at lower
% $Q^2$ is attributed to the omission of meson cloud effects. 

The collaboration thanks the Hall C technical staff and the Jefferson
Lab Accelerator Division for their outstanding support during the
experiment. This work was supported in part by the U.S. Department of Energy, the U.S.
National Science Foundation, the
Italian Institute for Nuclear Research, the French Commissariat
\`a l'Energie Atomique and Centre National de la Recherche Scientifique
(CNRS), and the Natural Sciences and Engineering
Research Council of Canada. Authored by Jefferson Science Associates, LLC under U.S. DOE Contract No. DE-AC05-06OR23177.

\bibliography{main}

\end{document}